\documentclass[pamm,a4paper,fleqn]{w-art}
\usepackage{times,cite,w-thm}
\usepackage[T1]{fontenc}
\usepackage[utf8]{inputenc}
\usepackage{amsfonts}
\usepackage{graphicx}
\begin{document}

\TitleLanguage[EN]
\title[Improved time integration for phase-field crystal models of solidification]{Improved time integration for phase-field crystal models of solidification}

\author{\firstname{Maik} \lastname{Punke}\inst{1,}%
\footnote{Corresponding author: e-mail \ElectronicMail{maik.punke@tu-dresden.de}, 
     phone +49\,351\,463-41202,
     fax +49\,351\,463-37096}}

\author{\firstname{Steven M.} \lastname{Wise}\inst{2}}%
\author{\firstname{Axel} \lastname{Voigt}\inst{1,3}}%
\author{\firstname{Marco} \lastname{Salvalaglio}\inst{1,3}%
    }
    
\address[\inst{1}]{\CountryCode[DE]Institute of Scientific Computing, TU Dresden, 01062 Dresden, Germany}
\address[\inst{2}]{\CountryCode[US]Department of Mathematics, The University of Tennessee, Knoxville, TN 37996, USA}
\address[\inst{3}]{\CountryCode[DE]Dresden Center for Computational Materials Science, TU Dresden, 01062 Dresden, Germany}

\AbstractLanguage[EN]
\begin{abstract}
We optimize a numerical time-stabilization routine for the phase-field crystal (PFC) models of solidification. By numerical experiments, we showcase that our approach can improve the accuracy of underlying time integration schemes by a few orders of magnitude. We investigate different time integration schemes. Moreover, as a prototypical example for applications, we extend our numerical approach to a PFC model of solidification with an explicit temperature coupling.
\end{abstract}
\maketitle                   

\section{Introduction}
The so-called phase-field crystal (PFC) model \cite{Elder2002,Elder2004,Emmerich2012}
is a prominent approach to describe crystal structures at large (diffusive) timescales through a continuous, periodic order parameter representing the atomic density. The PFC model is based on a Swift-Hohenberg free energy functional \cite{Elder2002,Elder2004,Emmerich2012}, which can be written as
\begin{equation}
F\left[\psi\right] = \int_\Omega \dfrac{\lambda}{2}\psi^2 -  \dfrac{\psi^3}{6} + \dfrac{\psi^4}{12}+ \dfrac{\kappa}{2}\left(-2 \left| \nabla \psi \right| ^2 + \left(\nabla^2 \psi \right)^2 \right) \rm{d}\mathbf{r}.   
\label{eq:F1}
\end{equation}
The scalar order parameter $\psi(\mathbf{r},t)$ \footnote{dependence on space and time are omitted elsewhere for the sake of readability, the same applies for other quantities introduced in the following.} is related to the atomic number density, and $\Omega \in \mathbb{R}^n$ ($n=2$ in this work). A set of parameters $\lambda,\,\kappa > 0$ characterize the phase space and material properties. Together with appropriate boundary- and initial conditions, the dynamics of $\psi$ is described via a conservative
($\mathrm{H}^{-1}$)
gradient flow of $F$. The time evolution reads
\begin{equation}
\begin{split}
\partial _t \psi &= \nabla^2 \dfrac{\delta F\left[\psi\right]}{\delta \psi}= L[\psi] + N[\psi],
\end{split}
\label{eq:pfc}
\end{equation}
with $L[\psi]$ and $N[\psi]$ a linear and nonlinear differential operator, respectively:
\begin{equation}
\begin{split}
L[\psi] &=  \left(\lambda \nabla^2+ 2\kappa \nabla^4 + \kappa \nabla^6\right) \psi,\\
N[\psi] &= \nabla^2 \left(-\dfrac{\psi^2}{2}+\dfrac{\psi^3}{3} \right).
\end{split}
\label{eq:pfc2}
\end{equation}
\begin{figure}[t]
	\centering
	\includegraphics{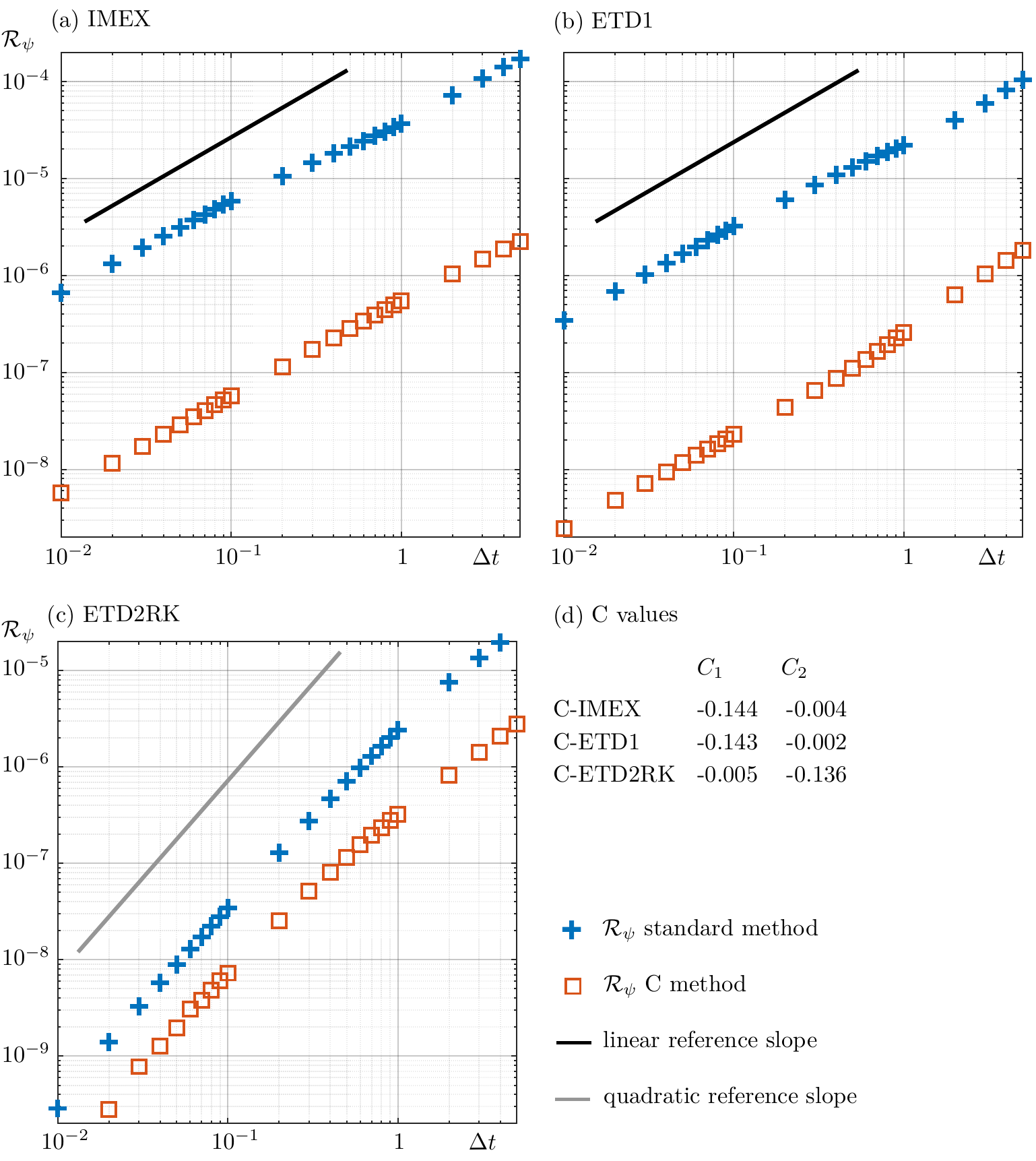}
	\caption{Convergence study of standard time stepping schemes ($C_1=C_2=0$) compared to their improved versions for equation~\eqref{eq:pfc} with a Fourier pseudo spectral space discretization and a simulation setup similar to Figure~\ref{fig:crystal}. (a) (C-)IMEX scheme (b) (C)-ETD1 scheme (c) (C-)ETD2RK scheme (d) corresponding $C_1$, $C_2$ values.} 
	\label{fig:pfc_convStudy}
\end{figure}

Numerical solutions of the partial differential equation \eqref{eq:pfc} can be computed using a Fourier pseudo-spectral method for spatial discretization, enforcing periodic boundary conditions in combination with a suitable time discretization \cite{baskaranNumeric,dongNumeric,warrenNumeric,ChengNumeric}. As time stepping schemes, we consider a linear first-order semi-implicit (IMEX) scheme, which treats $L[\psi]$ fully implicitly and $N[\psi]$ fully explicitly, a first-order exponential time integration method (ETD1), and a second-order exponential time integration method of Runge Kutta type (ETD2RK) \cite{cox2002exponential}. 

Additionally, we exploit the numerical time-stabilization routine presented in \cite{elsey2013simple}. In combination with an underlying time stepping scheme, this approach features a convex-concave splitting of the free energy controlled by parameters $C_1$ and $C_2$:
\begin{equation}
\begin{split}
L &\mapsto \widetilde{L} =  L + C_1 \nabla^2 - C_2 \nabla^4,\\
N &\mapsto \widetilde{N} = N - C_1 \nabla^2 + C_2 \nabla^4.
\end{split}
\label{eq:operators}
\end{equation}
We show in the following that they can be tuned to achieve an improved numerical accuracy, and that the best performances are here obtained for negative values.
For every time integration scheme, we determine optimal $C_1$, $C_2$ by minimizing the difference of the error (least square of the difference) in the free energy decay with respect to a numerical reference solution and denote the resulting time stepping schemes by C-IMEX, C-ETD1, and C-ETD2RK. 

\begin{figure}[b]
	\centering
	\includegraphics{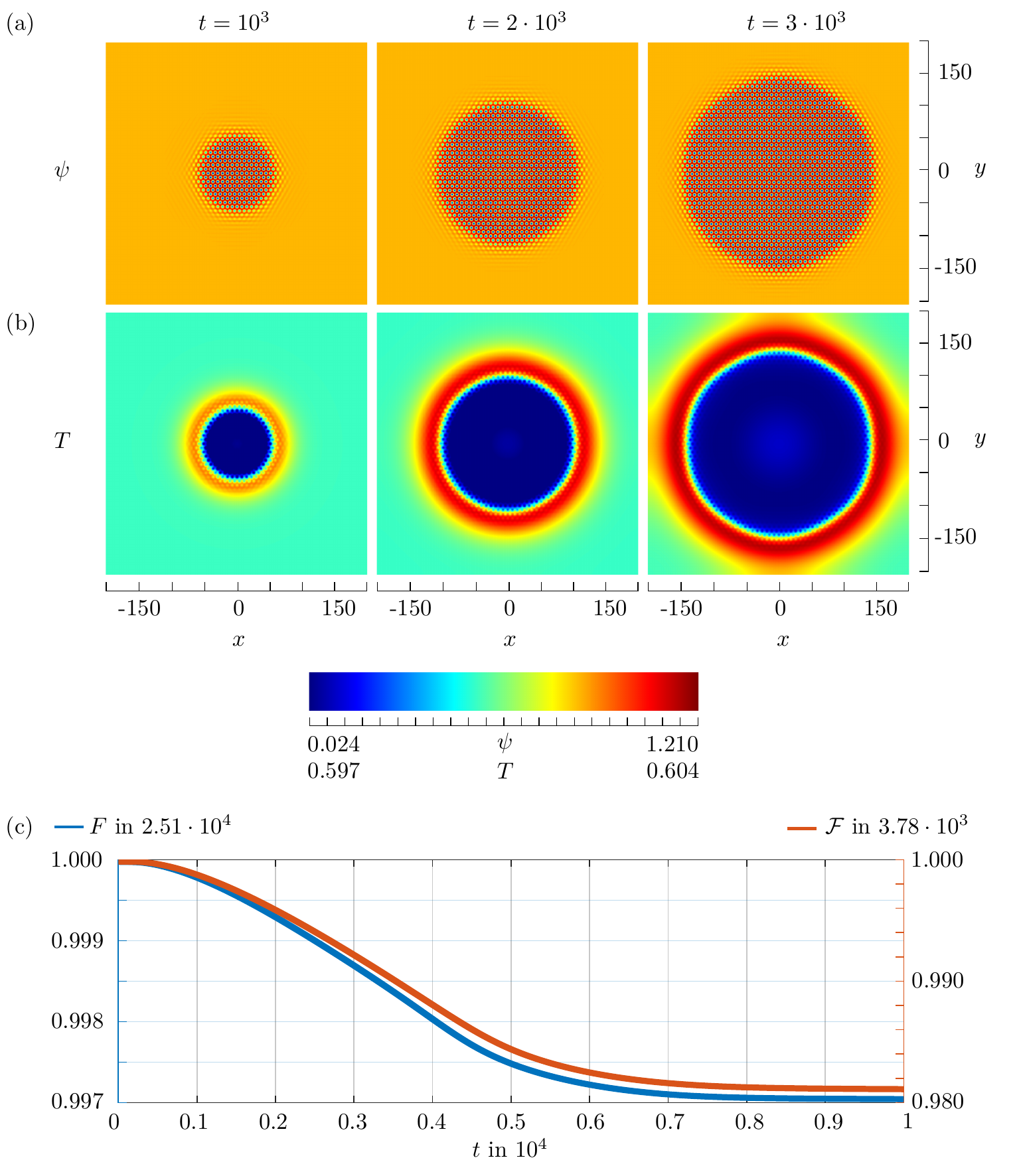}
	\caption{Numerical solution of equation~\eqref{eq:pfc} and \eqref{eq:psiEvol} for a given set of parameters ($\lambda=0.6$, $\kappa=0.45981$, $A=0.849$, $\gamma=0.06$, $\vartheta=0.1$, $T_0=0.6$, $M=0.1$) and $\Omega=[-200,\, 200]^2$. (a) density fields $\psi$ during crystal growth at $t=10^3$, $t=3\cdot 10^3$ and $t=10^4$ for equation~\eqref{eq:pfc} and \eqref{eq:psiEvol} are shown (b) temperature fields $T$ during crystal growth at $t=10^3$, $t=3\cdot 10^3$ and $t=10^4$ for equation~\eqref{eq:psiEvol} are shown (c) time evolution of the free energies $F$ and $\mathcal{F}_\beta$. 
	}
	\label{fig:crystal}
\end{figure}
\section{Numerical parameter study}
A convergence study of the considered numerical methods is reported in Figure~\ref{fig:pfc_convStudy}, for a specific set of model parameters leading to the growth of a crystal in a domain $\Omega= [-200, 200]^2$ (initial condition as in Figure~\ref{fig:crystal}~(a)). The numerical simulations are performed on an uniform grid with an element size of $\rm{d}x = \rm{d}y = 0.78125$ and for time steps $\Delta t \in [10^{-2}, 5]$. Figure~\ref{fig:pfc_convStudy} shows the residual of $\psi$, $\mathcal{R}_\psi$, evaluated as the discrete $L_2$ distance from a numerical reference solution for different $\Delta t$. The schemes IMEX, ETD1 and ETD2RK ($C_1=C_2=0$) and the corresponding ones including the splitting \eqref{eq:operators}, \mbox{C-IMEX}, \mbox{C-ETD1}, and \mbox{C-ETD2RK} are compared, with a reference solution corresponding to the C-ETD2RK scheme with $\Delta t =0.01$. As expected, the \mbox{(C-)IMEX} and \mbox{(C-)ETD1} schemes converge linearly, and the \mbox{(C-)ETD2RK} scheme converges quadratically for a decreasing time step size $\Delta t$. For a fixed $\Delta t$, the \mbox{C-IMEX} and \mbox{C-ETD1} schemes give two orders of magnitude smaller errors than the IMEX and ETD1 schemes, respectively, whereas the \mbox{C-ETD2RK} scheme gives a half order of magnitude smaller error than the ETD2RK scheme. By fixing an accuracy instead, the \mbox{C-IMEX} and \mbox{C-ETD1} scheme allow for two orders of magnitude larger timesteps than the IMEX and ETD1 scheme, respectively, whereas the \mbox{C-ETD2RK} scheme allows for two times larger timesteps than the ETD2RK scheme. Note, that the  \mbox{(C)-IMEX} and \mbox{(C)-ETD1} integration schemes show similar computational costs for a time step update. Due to an additional intermediate step, the time step updates with the \mbox{(C)-ETD2RK} integration schemes are twice as expensive as with the \mbox{(C)-IMEX} or \mbox{(C)-ETD1} approaches. 

\section{Extension to an explicit temperature coupling}

\begin{figure}[h]
	\centering
	\includegraphics{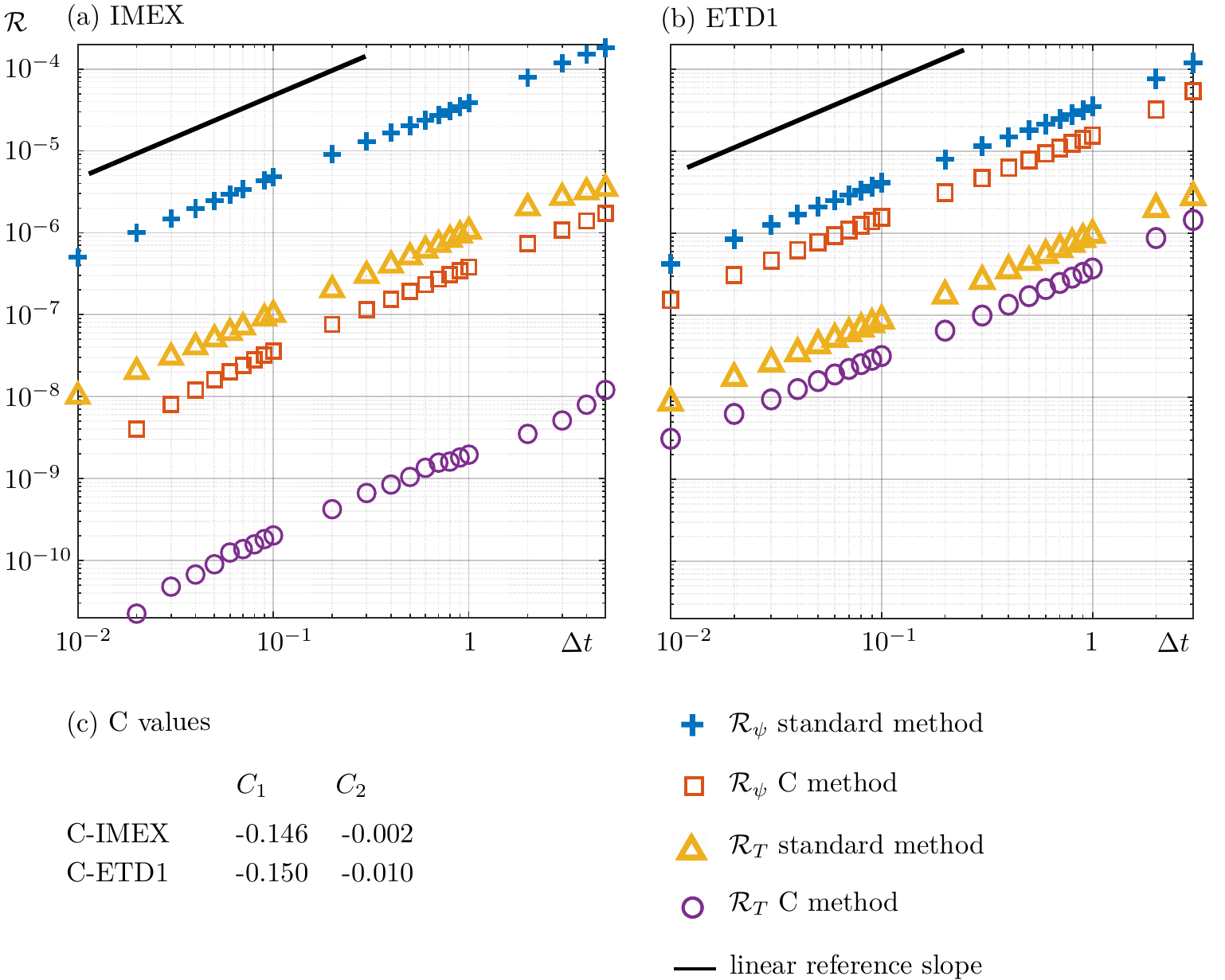}
		\caption{Convergence study of standard time stepping schemes ($C_1=C_2=0$) compared to their improved versions for equation~\eqref{eq:psiEvol} with a Fourier pseudo spectral space discretization and a simulation setup from Figure~\ref{fig:crystal}. The residuals of $\psi$ and $T$, $\mathcal{R}_\psi$, and $\mathcal{R}_T$ are shown. (a) (C-)IMEX scheme (b) (C)-ETD1 scheme (c) corresponding $C_1$, $C_2$ values.} 
	\label{fig:convStudy_T}
\end{figure}

So far, a basic formulation of the PFC model has been considered. However, many extensions have been proposed involving coupling of equation~\eqref{eq:pfc} with additional equations. For instance, an extension of the classical PFC model \eqref{eq:F1} including heat transfer through a temperature field has been proposed in \cite{wang2021thermodynamically,Punke_2022}. Therein, the (dimensionless) Helmholtz free energy functional reads 
\begin{equation}
    \begin{split}
\mathcal{F}\left[\psi, T\right] & := F\left[\psi\right] - \int_\Omega \vartheta \mathrm{ln}(T) + \frac{1}{T}\gamma (\psi +1) \rm{d}\mathbf{r},
\label{eq:F2}
    \end{split}
\end{equation}
where $T(\mathbf{r},t)$ is the dimensionless temperature with $T=1$ at the melting point, and $\vartheta, \gamma >0$ are additional parameters.  The dynamics are then given by the coupled evolution of $T$ and $\psi$, reading 
\begin{equation}
\begin{split}
   \vartheta \partial_t T- \gamma \partial _t \psi    &= M \nabla^2 T, \\
    \partial_t \psi &= \nabla^2  \dfrac{\delta \mathcal{F}\left[\psi,T\right]}{\delta \psi},
    \end{split}
    \label{eq:psiEvol}
\end{equation}
with $M>0$ a parameter corresponding to the thermal diffusivity assumed to be constant. We consider additional initial and boundary conditions for $T$. In \cite{Punke_2022}, similar concepts to those illustrated in the previous sections have been exploited and are here investigated further. Figure~\ref{fig:crystal}~(a)-(c) shows the density and temperature fields of a growing crystal at intermediate time steps with the corresponding free energy evolution. 
Similar to equation~\eqref{eq:operators} we introduce $C_1$, $C_2$ and report a convergence study of the resulting \mbox{(C-)IMEX} and \mbox{(C-)ETD1} approaches, in which we compare the residuals of $\psi$ and $T$, $\mathcal{R}_\psi$ and $\mathcal{R}_T$, respectively, cf. Figure~\ref{fig:convStudy_T}. The residuals are evaluated as the discrete $L_2$ distance from a numerical reference solution, corresponding to a \mbox{C-IMEX} scheme with $\Delta t = 0.01$.  As expected, all schemes converge linearly for a decreasing time step size $\Delta t$. Moreover, for a fixed $\Delta t$, the \mbox{C-IMEX} approach gives two orders of magnitude smaller error than the IMEX approach for $\psi$ and almost three orders of magnitude smaller error for $T$. By fixing an accuracy instead, the \mbox{C-IMEX} approach allows for two orders of magnitude larger timesteps than the IMEX approach, resulting in a further improvement compared to the performances achieved for the numerical integrations as reported in \cite{Punke_2022}. For the \mbox{C-ETD1} approach only small performance improvements compared to the \mbox{ETD1} approach are achieved.

\section{Conclusion}

We studied the numerical integration of the governing equations in the PFC model with different time discretizations and featuring a numerical time-stabilization \cite{elsey2013simple}. Also, we showcased by numerical examples that this approach works for both first-order and second-order exponential time integration methods. We outlined the main criteria to determine numerical parameters and investigate the performances for a prototypical simulation, which may be exploited as a benchmark. The applicability to extended PFC models featuring the coupling with additional variables is shown for a model coupling Eq.~\eqref{eq:pfc} with heat flow, as reported in \cite{Punke_2022}. A \mbox{C-IMEX} and a \mbox{C-ETD1} integration scheme are considered. Especially for the \mbox{C-IMEX} integration scheme, a further improvement is here achieved with respect to a similar approach exploited in \cite{Punke_2022} thanks to an extended parametrization of the splitting \eqref{eq:operators}.

\begin{acknowledgement}
MP and MS acknowledge support from the German Research Foundation (DFG) under Grant No.~SA4032/2-1. AV acknowledges support from the German Research Foundation (DFG) within SPP1959 under Grant No.~Vo899/20-2. SMW gratefully acknowledges support from the US National Science Foundation under grant NSF-DMS 2012634. Computing resources have been provided by the Center for Information Services and High-Performance Computing (ZIH) at TU Dresden, and by J\"ulich Supercomputing Center under grant PFAMDIS.
\end{acknowledgement}

\vspace{\baselineskip}

\end{document}